\documentclass[twocolumn,showpacs,preprintnumbers,amsmath,amssymb,superscriptaddress,prl]{revtex4}
\usepackage{graphicx}% Include figure files
\usepackage{dcolumn}% Align table columns on decimal point
\usepackage{bm}% bold math
\usepackage{subfigure}
\usepackage{color}
\usepackage{amsfonts}%
\usepackage{amssymb}
\usepackage{amsmath}%
\begin{document}

%\preprint{APS/123-QED}

\title{Origin of the anomalous Hall Effect in overdoped n-type cuprates:\\ current vertex corrections due to antiferromagnetic fluctuations}

\author{G. S. Jenkins}
\author{D. C. Schmadel}%
\author{P. L. Bach}%
\author{R. L. Greene}%
\affiliation{%
Center for Nanophysics and Advanced Materials\\Department of
Physics, University of Maryland, College Park, Maryland 20742, USA
}%

\author{X.~B\'echamp-Lagani\`{e}re}
\author{G. Roberge}
\author{P. Fournier}
\affiliation{Regroupement qu\'eb\'ecois sur les mat\'eriaux de
pointe, D\'epartement de Physique, Universit\'e de Sherbrooke,
Sherbrooke, Qu\'ebec, Canada, J1K 2R1}
\author{Hiroshi Kontani}
\affiliation{
Department of Physics, Nagoya University, Furo-cho, Nagoya 464-8602, Japan% with \\
}%

\author{H. D. Drew}
\affiliation{%
Center for Nanophysics and Advanced Materials\\Department of
Physics, University of Maryland, College Park, Maryland 20742, USA
}%
\date{\today}% It is always \today, today,
             %  but any date may be explicitly specified

\begin{abstract}

The anomalous magneto-transport properties in electron doped
(n-type) cuprates were investigated using Hall measurements at THz
frequencies.  The complex Hall angle was measured in overdoped
Pr$_{\rm 2-x}$Ce$_{\rm x}$CuO$_{\rm 4}$ samples (x=0.17 and 0.18)
as a continuous function of temperature above $T_c$ at excitation
energies 5.24 and 10.5 meV.  The results, extrapolated to low
temperatures, show that inelastic scattering introduces
electron-like contributions to the Hall response.  First principle calculations of the Hall angle 
that include current vertex corrections (CVC)  induced by electron 
interactions mediated by magnetic fluctuations in the Hall conductivity 
reproduce the temperature, frequency, and doping dependence 
of the experimental data. These results show
that CVC effects are the source of the anomalous Hall transport
properties in overdoped n$\text{-}$type cuprates.

%Valid PACS numbers may be entered using the \verb+\pacs{#1}+ command.
\end{abstract}

% PACS, the Physics and Astronomy
                             % Classification Scheme.
%\keywords{Suggested keywords}%Use showkeys class option if keyword
                              %display desired
\maketitle

%\section{%\label{sec1:level1}Introduction:\protect\\}
The anomalous properties of the cuprates above the superconducting transition temperature have presented many puzzling challenges to the paradigms of condensed matter physics.  In particular the unusual magnetotransport has often been cited as evidence that the cuprates are not Fermi liquids
\cite{AndersonBook, Kontani_2008Review}.  Despite the large simple convex hole-like Fermi surfaces observed by angular resolved photoemission (ARPES), both optimally doped n- and p- type cuprates exhibit anomalous and strongly temperature dependent Hall coefficients.  In overdoped n-type cuprates, $R_{\rm H}$ has zero crossings that suggest a mixed electron and hole response. Consensus on an explanation for the apparent non-Fermi liquid behavior of the cuprates has not been achieved despite much theoretical and experimental effort \cite{AndersonBook,Kontani_2008Review}.

One proposed explanation involves current vertex corrections (CVC) to 
the standard relaxation time approximation (RTA) for the Hall 
conductivity \cite{Kontani_2008Review}.  
In this Fermi-liquid scenario, inelastic electron interactions mediated by
antiferromagnetic (AF) fluctuations are the operative mechanism.  In an alternative explanation thermally induced magnetic fluctuations at finite temperatures reconstruct the FS dynamically leading to fluctuating electron and hole-like Fermi surface
segments over an area determined by the AF correlation length and
a time scale associated with the AF correlation time \cite{HarrisonPRL2007}. In both of these scenarios, Fermi liquid behavior is recovered at T$=$0 where inelastic scattering and thermal fluctuations vanish.

In this letter we directly address these issues
by extending magnetotransport measurements into
the frequency domain in the electron doped cuprates. We report temperature and doping
dependent Hall data on overdoped Pr$_{\rm 2-x}$Ce$_{\rm x}$CuO$_{\rm 4}$ (PCCO) at THz
frequencies $\lesssim\mspace{-3.0mu}\text{10~meV}$. 
At finite frequencies near zero temperatures where inelastic scattering is still operative but magnetic fluctuations are weak, we observe a strong suppression of the Hall response which corresponds to electronlike contributions coming into $\sigma_{\rm xy}$, even at T$=$0. Our experimental observations
together with direct comparisons between our magnetotransport data
and first-principle calculations of the frequency dependent Hall angle
which include CVCs offer compelling evidence in support of the Fermi
liquid interpretation.

%%%%%%%%%%%%%%%%%%%%%%%%%%%%trim=l b r t
\begin{figure}[!t]
\includegraphics[scale=.45,clip=true, trim = 50 0 180 15]{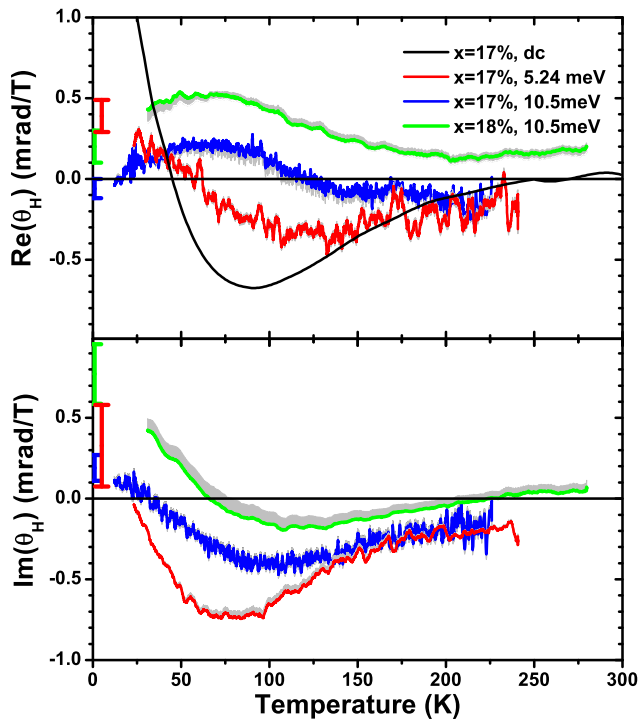}% Here is how to import EPS art
\caption{\label{fig:ReandImHA}(color online) The real and imaginary parts of the
Hall angle measured at dc, 5.24, and 10.5 meV. The negative value
at finite temperature in both the real and imaginary parts of the
Hall angle signify strong electron-like contributions which
dominate the $\sigma_{\rm xy}$ response, a clear deviation from
the simple hole-like Fermi surface observed by ARPES. The brackets
at low temperature are the zero temperature extrapolations of the
data as described in the text. Depicted in grey with each data set are
error bars derived from the juxtaposition of uncertainties in
thickness, $\sigma_{\rm xx}$, and $\theta_{\rm F}$. The included
dc Hall data is taken from Ref. \cite{Dagan_thetaHvsDoping}.}
\end{figure}
%%%%%%%%%%%%%%%%%%%%%%%%%%%

%%%%%%%%%%%%%%%%%%%%%%%%%%%%%%%%%
\begin{figure}[t]
\includegraphics[width=.8\linewidth]{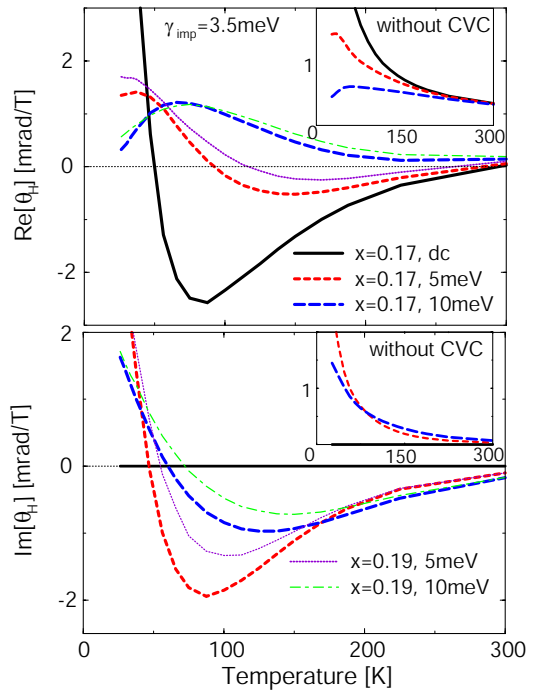}
\caption{(color online) Prediction of the real and imaginary part of the Hall
angle by incorporating current-vertex corrections (CVCs) induced by
magnon scattering. The Hall angle calculated in the FLEX approximation without including CVCs is shown in the insets.} \label{fig:FLEX}
\end{figure}
%%%%%%%%%%%%%%%%%%%%%%%%%%%%%%%%%

Thin film Pr$_{\rm 2-x}$Ce$_{\rm x}$CuO$_{\rm 4}$ c-axis oriented
samples were grown via pulsed laser deposition onto 100 $\mu$m
thick $\text{LaSrGaO$_{\rm 4}$}$ (001) substrates. The two samples
reported here have a chemical doping of x$=$0.17
\cite{PatrickXtalGrowth} and 0.18 with thicknesses 40 and 125 nm
and $T_c$'s of 13 and 9 K, respectively. Resistivity and dc Hall measurements were similar to previously reported
data \cite{QCP_16percent,Dagan_thetaHvsDoping}. However, the low temperature dc Hall coefficient $R_{\rm H}$ measured on these
new films is consistent with the simple large
hole-like Fermi surface centered at ($\pi$,$\pi$) measured by
ARPES \cite{ARPES_edoped_Matsui}. Luttinger's thereom is satisfied
since the carrier number density associated with the FS volume
measured by ARPES is consistent with the stoichiometric doping
level \cite{QCP_16percent,Lin_Millis,Footnote_BetterXtals}. As temperature is raised, the dc Hall coefficient rapidly decreases and becomes negative at a
temperature which increases with $x$, and $R_{\rm H}$ eventually
returns to positive values around room temperature. 

Below a doping of $\sim$16\% where PCCO undergoes a quantum phase transformation to an antiferromagnetic state, the low temperature dc Hall coefficient deviates sharply from the observed overdoped behavior \cite{QCP_16percent}.  At and below the critical doping, ARPES data show the fractionalization of the Fermi surface into Fermi
arcs \cite{ARPES_edoped_Matsui,ARPES_edoped_Armitage} consistent with recent quantum oscillations experiments \cite{NCCO_QO}.  In this letter we address the overdoped paramagnetic phase in which the large holelike Fermi surface remains intact. 

The Faraday rotation and circular dichroism were measured
(expressed as the complex Faraday angle, $\theta_{\rm F}$) at
discrete frequencies as a continuous function of temperature. The
output of a far-infrared molecular vapor laser was polarization
modulated with a rotating quartz quarter-wave plate and
subsequently transmitted through the c-axis oriented sample at
normal incidence in an applied magnetic field up to 8T. The
detector signal was harmonically analyzed to directly extract 
the complex Faraday angle, a technique that is detailed elsewhere
\cite{Jenkins2009PRB,GJThesis}.

In the thin film limit, the complex Hall angle is related to the
Faraday angle via $\theta_{\rm H} = (1+ \frac{n+1}{Z_0 \sigma_{\rm
xx} d}) \theta_{\rm F}$, where $\sigma_{\rm xx}$ is the
longitudinal conductivity, $n$ is the index of refraction of the
substrate, $Z_0$ is the impedance of free space, and $d$ is the
thickness of the film \cite{GJThesis}. FTIR-spectroscopic transmission
measurements were performed in the spectral range from 2 to 13 meV
at a set of discrete temperatures ranging from 5 to 300 K, and the
complex conductivity $\sigma_{\rm xx}$ was extracted by fitting to
a Drude form.

We have measured the real and imaginary parts of the Hall angle at
10.5 and 5.24 meV. The Hall signals were found to be linear in
field over the measured magnetic field range of $\pm$8 T.  The
complex Hall angle as a function of temperature is plotted in
Fig.\thinspace\ref{fig:ReandImHA}. As in the case of the dc Hall
effect in overdoped PCCO \cite{Dagan_thetaHvsDoping}, we observe a
complex temperature dependence with zero crossings that depend on
doping.

The observed Hall angle differs in important ways compared to the dc behavior (see Fig.\thinspace\ref{fig:ReandImHA}). Most importantly, Im($\theta_{\rm H}$) 
becomes nonzero. Im($\theta_{\rm H}$) is negative at high temperature crossing zero at a temperature that increases with
doping and frequency. The low
temperature Im($\theta_{\rm H}$) values are positive as expected for a holelike Fermi surface as observed by ARPES.  However the negative values at 
high temperatures is inconsistent with the holelike Fermi surface within Drude-like models.
The differences between the Re($\theta_{\rm H}$) 
and the dc Hall angle increase with frequency and doping.  The peak 
in Re($\theta_{\rm H}$) is expected in the simple Drude model for $\omega \sim \gamma_{\rm H}$.

Motivated by the simple
Fermi liquid-like behavior of the low temperature dc Hall
coefficient, we begin our discussion by comparing the infrared (IR) Hall response with a simple Drude
model where the complex Hall
angle in the weak field approximation is given by:
\begin{eqnarray}
\theta_{\rm H} = \frac{\sigma_{\rm xy}}{\sigma_{\rm xx}} =
\frac{\omega_{\rm H}}{\gamma_{\rm H} -
\it{i}\omega}\label{eq:HAdefinition}
\end{eqnarray} where $\sigma_{\rm xy}$ is the Hall conductivity, $\sigma_{\rm xx}$\ is
the longitudinal conductivity, $\omega_{\rm H}=q B / m_{\rm H}$ is
the Hall frequency,  $m_{\rm H}$ is the effective Hall mass, $\omega$ is the radiation frequency, 
$\gamma_{\rm H}$ is the Hall scattering
rate, B is the applied magnetic field,
and q is the effective charge of the quasi-particle. While Eq.\thinspace(\ref{eq:HAdefinition}) can not be expected to
accurately represent either the dc or IR Hall data of
Fig.\thinspace\ref{fig:ReandImHA} at finite temperature because of
the apparent combined electron- and hole-like response, it may be expected to correctly describe the low temperature results. 

To accomplish this it is necessary to extrapolate the Hall data to T$\simeq$0.  Generally in conducting condensed matter systems when the energy scale associated with temperature becomes less than the frequency, the temperature dependent conductivity rolls over to a flatter response. The exponent which may describe the low temperature finite frequency power law response is expected to be larger than that of the dc response. Therefore, we use as a guide the reported dc Hall coefficient, Hall angle, and longitudinal resistivity data \cite{QCP_16percent, Dagan_thetaHvsDoping} for 17\% and 18\% PCCO which were fit between 0 and 50 K and demonstrate temperature power laws with exponents ranging from 0.8 to 1.5. The zero temperature extrapolations of similar power law fits to the IR Hall angle data between 0 and 50 K with exponents ranging from 0.8 to 2 are shown in Fig.\thinspace\ref{fig:ReandImHA}. Even with these generous overestimates of uncertainties associated with the extrapolations, the IR Hall angle is clearly much smaller than the dc Hall angle (2.6 mrad/T and 3.4 mrad/T for the 17\% and 18\% doped samples, respectively) \cite{QCP_16percent, Dagan_thetaHvsDoping}.

This suppression of the finite frequency Hall response is more quantitatively characterized by comparing the low temperature Hall frequency $\omega_{\rm H}$ in the Drude representation to that expected from ARPES measurements. As noted previously, the low temperature dc Hall coefficient above the quantum critical doping is correctly given by the Fermi surface properties measured by ARPES. All the magnetotransport response functions including $\omega_{\rm H}$ can be derived directly from ARPES data within the RTA in which $\sigma_{\rm xx}$ and $\sigma_{\rm xy}$ are both expressed as integrals around the Fermi surface involving the Fermi velocities and scattering rates  \cite{Yakovenko-and-Drew, DrewMillisSigmaXX}. ARPES measurements determine the size and shape of the Fermi surface, Fermi velocities, and momentum distribution widths.  For optimal doping, the Hall frequency derived from the large holelike Fermi surface observed by ARPES \cite{ARPES_edoped_Matsui, Jenkins2009PRB} yields $\omega_{\rm H}^0$=.052 meV/T, equivalent to an effective Hall mass $m_{\rm H}$ = 2.2 $m_{\rm e}$. 

The estimated T$\simeq$0 Hall frequencies normalized to the values deduced from ARPES data, given by $\omega_{\rm H}/|\omega_{\rm H}^0|  =
- (\omega/|\omega_{\rm H}^0|)\thinspace [Im(1/\theta_{\rm
H})]^{-1}$,  are presented in Table\thinspace\ref{tab:tablea}. These data replicate our main observation that the Hall response is severely suppressed at finite frequency compared to the low temperature dc values and, equivalently, the RTA expectations deduced from ARPES.  The uncertainties associated with the IR Hall angle extrapolations are small compared to this suppression. Also discernable in Table\thinspace\ref{tab:tablea} is a rapid reduction in the Hall response with frequency similar to the rapid decrease in $R_{\rm H}$ and the dc Hall angle
\cite{Dagan_thetaHvsDoping} with temperature. This reduction with increasing frequency can be seen directly in the Table\thinspace\ref{tab:tablea} (rows 1 and 2), and both values are much smaller than the dc values deduced from ARPES measurements. It should also be noted that the reduction of the Hall response with temperature or frequency, expressed in the same units, is of a similar magnitude. From the doping dependence in Table\thinspace\ref{tab:tablea} (rows 2 and 3), it is also evident that the Hall frequency suppression increases as the doping is reduced from 18\% to 17\% toward the quantum critical doping.

%%%%%%%%%%%%%%%%%%%%%%%%%%%%%%%%%%%%%%%%%%%%
\begin{table}[t]
\begin{ruledtabular}
\begin{tabular}{cccc}
doping, x (\%) &$\omega$ (meV) &$\omega_{\rm H} / \omega_{\rm H}^0$\\
\hline
17  &5.24    &0.20$\pm$0.13 \\
17  &10.5   &0.04$\pm$0.02\\
18  &10.5    &0.16$\pm$0.04 \\
\end{tabular}
\end{ruledtabular}
\caption{\label{tab:tablea}Utilizing the extrapolated zero temperature complex Hall angle values in Fig.\thinspace\ref{fig:ReandImHA}, the
finite frequency zero temperature Hall frequency $\omega_{\rm H}$
may be extracted normalized to
the Hall frequency predicted by ARPES
measurements and Boltzmann transport theory,
$\omega_{\rm H}^0$=0.052 meV/T.}
\end{table}
%%%%%%%%%%%%%%%%%%%%%%%%%%%%%%%%%%%%%

The  observed reduction in the Hall angle with frequency (and/or temperature) 
corresponds to a
large suppression in $\sigma_{\rm xy}$ caused by the presence of
electronlike contributions. The electronlike contributions at T$\simeq$0 and finite frequency 
cannot be due to antiferromagnetic (AF) fluctuations since thermally excited AF fluctuations 
vanish at zero temperature. Instead these results suggest a breakdown 
of the RTA except when both temperature and frequency are zero where inelastic 
scattering vanishes. While the experiments implicate inelastic scattering as 
the underlying cause of the anomalous Hall effect in overdoped PCCO, the mechanism is not obvious from the experimental findings.  However, current vertex corrections to the conductivity in the presence of antiferromagnetic fluctuations can lead to currents 
that are not necessarily parallel to the quasiparticle velocity which can produce negative contributions to the Hall conductivity \cite{Kontani_2008Review}. In these nearly antiferromagnetic metals,
the Hall conductivity is strongly modified by the CVCs which represents electron-electron (Umklapp) scattering processes associated with the magnetic Brillouin zone \cite{Kontani_2008Review}.  Including the CVCs has successfully reproduced the strong  temperature dependence of dc Hall coefficient
in both electron- and hole-doped cuprates \cite{Kontani_2008Review}. These effects also modify the frequency dependent response since inelastic scattering is frequency dependent and occurs even at T$\simeq$0.

%%%%%%%%%%%%%%%%%%%%%%%%%%%%%%%%%%%%%%%%%%%%%%%%%%%%%%%%%%%%%

\begin{figure}[t]
\includegraphics[scale=.45,clip=true, trim = 0 0 0 0]{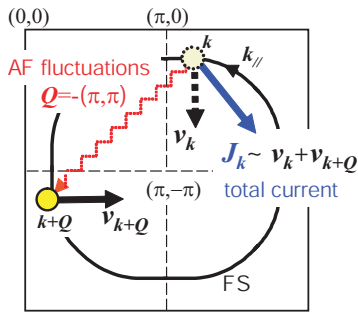}
\caption{(color online) Electron scattering process ${\bf k}\rightarrow {\bf
k+Q}$ due to AF fluctuations ${\bf Q}\approx (\pm\pi,\pm\pi)$. The
total current ${\bf J}_{\bf k}$ is composed of quasiparticle
velocities at ${\bf k}$ and ${\bf k+Q}$.
%The negative curvature of the effective Fermi surface
%gives rise to the negative Hall coefficient in PCCO.
} \label{fig:CVC}
\end{figure}

To examine the role of CVCs on the IR Hall response, we have calculated the frequency and temperature dependent conductivity for overdoped PCCO within   
the fluctuation-exchange (FLEX) conserving
approximation.  The starting point is the Hubbard model with model parameters
for the tight binding hopping amplitudes representing the band
structure ($t,t',t''$) and the Hubbard $U$
\cite{Kontani_2008Review}. In the FLEX approximation the effective
interaction potential is proportional to the derived dynamical
spin susceptibility which has the following functional form:
$\chi^s_{\bf q}(\epsilon) \propto \xi^2(1+\xi^2({\bf q}-{\bf
Q})^2-i\epsilon/\omega_{\rm sf})^{-1}$, where the AF correlation
length $\xi$ is much larger than the lattice spacing in PCCO at
low temperatures \cite{Motoyama}. With this interaction, an electron with ${\bf k}$
on the Fermi surface is scattered to ${\bf k+(-) Q}$ by absorbing
(or emitting) a low-energy AF magnon fluctuation.  This scattering
process in nearly AF metals not only shortens the quasiparticle
lifetime (i.e., hot-spot), but also changes the direction of
electron motion. In Fermi liquid theory, the former and the latter
effects are described by the self-energy and the CVC,
respectively.   How the electron scattering due to magnetic fluctuations leads to strong CVC effects is illustrated in     
Fig.\thinspace\ref{fig:CVC}.  An excited
electron at ${\bf k}$,  after a quasiparticle lifetime
$1/2\gamma_{\rm AF}$, is scattered to ${\bf k+Q}$ due to AF
fluctuations ${\bf Q}\approx (\pm\pi,\pm\pi)$. Therefore, ${\bf
J}_{\bf k}\propto {\bf v}_{\bf k}+ {\bf v}_{\bf k+Q}$ in the
hydrodynamic regime $\omega\ll\gamma_{\rm AF}$.

%%%%%%%%%%%%%%%%%%%%%%%%%%%%%%%%%
%%%%%%%%%%%%%%%%%%%%%%%%%%%%%%%%%

Figure \ref{fig:FLEX} shows the numerical results for the ac Hall angle obtained by the
FLEX+CVC approximation. The nearest neighbor hopping integral is
taken as $t=-0.36$ eV and the quasiparticle damping rate due to
elastic scattering was set at $\gamma_{\rm imp}=3.5$ meV
corresponding to $\rho_{\rm imp}\sim 5$ $\mu\Omega\thinspace {\rm cm}$. In
the absence of CVC the complex frequency dependent $\theta_{\rm
H}$ is positive due to the large hole-like Fermi surface 
corresponding to the ARPES data on electron overdoped cuprates.
In Fig.\thinspace\ref{fig:FLEX} in which CVCs are included, both
Re($\theta_{\rm H}$) and especially Im($\theta_{\rm H}$) take on
increasingly negative values as temperature is lowered from room
temperature and the AF fluctuations become stronger.
Below 100 K when the damping rate due to AF fluctuations
$\gamma_{\rm AF}$ becomes smaller than the elastic scattering
rate, CVCs become less important.   However, since $\gamma_{\rm AF}$ increases with $\omega$, CVC effects will be more important for larger $\omega$.  As $\omega$ and T approach zero, the
inelastic scattering reduces and both Re($\theta_{\rm H}$)
and Im($\theta_{\rm H}$) become positive. In the collisionless regime where $\omega\gg\gamma_{\rm AF}$ the electron-electron scattering becomes less important and ${\bf J}_{\bf k}\sim {\bf v}_{\bf k}$.

Also, as can be seen in
Fig.\thinspace\ref{fig:FLEX} as $T$ approaches zero,
Re($\theta_{\rm H}$) at 5 and 10 meV is much smaller than the zero
frequency value as observed directly.  In addition, the
figure shows the calculated results for two different doping
levels which demonstrate the reduction of the CVC effects as the
system is doped further from the AF phase boundary.

Although the temperature, frequency, and doping dependence of $\theta_{\rm H}$ is qualitatively well described by the FLEX+CVC theory, the overall amplitude of the experimental data in Fig.\thinspace\ref{fig:ReandImHA}  is smaller by a factor of 2 to 3 compared with the theoretical results in Fig.\thinspace \ref{fig:FLEX}. This discrepancy may result from the fact that the FLEX approximation, as a low energy theory, can provide an accurate description of the magnetic correlation effects but can not give a good account of the Mott correlations in the Hubbard model.  In particular, the theory fails to account for the high frequency optical transitions for $\omega \gtrsim U$.  This implies, by the optical conductivity sum rules, that the theory overestimates the weight of the low energy Drude-like response.  The suppression of the Drude-like spectral weight in the cuprates has been discussed by Millis \textit{et al.} \cite{Millis-sxxMott,DrewMillisSigmaXX}.  Experiments suggest that the Drude oscillator strength associated with $\sigma_{\rm xy}$  is more strongly suppressed than that associated with $\sigma_{\rm xx}$ \cite{SchmadelPRBRapid,AZimmersFilmGrowth}.  Therefore, since $\theta_{\rm H}=\sigma_{\rm xy}/\sigma_{\rm xx}$, there is an overall suppression of the Hall angle Drude-like oscillator strength \cite{Coleman-sumrules} that is not captured by the FLEX theory.  Further study of these Mott correlation effects on $\sigma_{\rm xy}$ are underway \cite{MillisPrivate}.

In conclusion, the low temperature Hall response of PCCO is observed to be strongly suppressed at finite frequencies.  The full complex temperature, frequency, and doping dependencies of $\theta_H$ in
overdoped PCCO are found to be naturally reproduced in the FLEX+CVC
approximation for this strongly interacting nearly
antiferromagnetic electron system. These results show that
current vertex corrections contribute significantly to the
magnetotransport properties of PCCO and, by extension, suggest  that 
CVC effects may also be the underlying mechanism causing the
anomalous Hall transport in the hole doped cuprate systems \cite{Kontani_2008Review}.

The authors wish to acknowledge the support of NSERC, FQRNT, CFI,
CNAM, and NSF (DMR-0653535 and DMR- 0303112). We thank S.
Pelletier and K. D. Truong for their technical assistance.

%\newpage %Just because of unusual number of tables stacked at end
\bibliography{Overdoped_PCCOv6}% Produces the bibliography via BibTeX.

\begin{thebibliography}{21}
\bibitem[{Foo()}]{Footnote_BetterXtals}
\eprint{This is in contrast with earlier dc Hall measurements on overdoped PCCO
  which deviated somewhat from the ARPES expectation
  \cite{QCP_16percent,Lin_Millis} which is attributed to improvements in the
  film growth process \cite{PatrickXtalGrowth}.}
\end{thebibliography}

\end{document}